\documentclass{llncs}
\usepackage{ltl}

\usepackage{tikz}
\usepackage{tikzscale}
\usepackage{pgfplots}
\pgfplotsset{
width=10cm,
compat=1.9}
\tikzset{every picture/.style=semithick,
}%every axis plot/.append style=thick}

\usetikzlibrary{positioning,mindmap,automata,fit,backgrounds,shapes, 
arrows,shapes.misc,patterns}
\usetikzlibrary{decorations.pathmorphing}
\usetikzlibrary{decorations.markings}

\usepackage{ltl}
\usepackage{tabulary}
\usepackage{url}
\usepackage{todonotes}
\usepackage{url}
\usepackage{cite}
\usepackage{xspace}
\usepackage{graphicx}
\usepackage{subfigure}
\usepackage{caption}
\usepackage{color}
\usepackage{multicol}
\usepackage{hyphenat}
\usepackage{fancyvrb}
\usepackage{lipsum}
\usepackage{stmaryrd}
\usepackage{wrapfig}

\usepackage[ruled,vlined,linesnumbered]{algorithm2e}
\SetKwInOut{Input}{Input}
\SetKwInOut{Output}{Output}
\SetKwProg{Fn}{Function}{}{}
\DontPrintSemicolon

\newcommand{\Paths}[2]{\mathit{Paths}^{#1}{(#2)}}
\newcommand{\fPaths}[2]{\mathit{Paths}^{#1}_{\mathit{fin}}{(#2)}}

\newcommand{\dtmc}{\mathcal{M}}
\newcommand{\LTL}{\textsf{\small LTL}\xspace}
\newcommand{\CTL}{\textsf{\small CTL}\xspace}
\newcommand{\CTLstar}{\textsf{\small CTL$^*$}\xspace}
\newcommand{\PCTL}{\textsf{\small PCTL}\xspace}
\newcommand{\PCTLstar}{\textsf{\small PCTL$^*$}\xspace}
\newcommand{\HyperPCTL}{\textsf{\small HyperPCTL}\xspace}
\newcommand{\HyperLTL}{\textsf{\small HyperLTL}\xspace}
\newcommand{\HyperCTLstar}{\textsf{\small HyperCTL$^*$}\xspace}

\renewcommand{\qed}{$~\blacksquare$}
\newcommand{\naturals}{\mathbb{N}_{>0}}
\newcommand{\naturalszero}{\mathbb{N}_{\geq 0}}

\newcommand{\AP}{\mathsf{AP}}
\newcommand{\Next}{\X}

\newcommand{\pr}{\mathbb{P}}
\renewcommand{\Pr}{\mathit{Pr}}

\newcommand{\tpm}{\mathbf{P}}
\newcommand{\tru}{\mathtt{true}}
\newcommand{\fals}{\mathtt{false}}
\newcommand{\quant}{\mathbb{Q}}

\newcommand{\dbsim}{\mathit{dbSim}}

\newcommand{\qout}{\mathit{qOut}}

\newcommand{\comp}[1]{\textsf{\small #1}}

\newcommand\donotshow[1]{}
\newcommand{\F}{\LTLdiamond}
\newcommand{\G}{\LTLsquare}
\newcommand{\U}{\,\mathcal U\,}
\newcommand{\X}{\LTLcircle}

\DefineVerbatimEnvironment{code}{Verbatim}{fontsize=\small}

\title{HyperPCTL: A Temporal Logic for Probabilistic Hyperproperties}

\author{Erika {\'A}brah{\'a}m\inst{1} \and Borzoo Bonakdarpour\inst{2}}

\institute{RWTH Aachen University, Germany \and Iowa State University, USA}

\begin{document}

\maketitle

\begin{abstract}

  In this paper, we propose a new logic for expressing and reasoning
  about probabilistic hyperproperties. \emph{Hyperproperties}
  characterize the relation between different independent executions
  of a system. \emph{Probabilistic} hyperproperties express
  quantitative dependencies between such executions. The standard
  temporal logics for probabilistic systems, i.e., \PCTL and \PCTLstar
  can refer only to a single path at a time and, hence, cannot express
  many probabilistic hyperproperties of interest. The logic proposed
  in this paper, \HyperPCTL, adds explicit and simultaneous
  quantification over multiple traces to \PCTL. Such quantification
  allows expressing probabilistic hyperproperties. A model checking
  algorithm for the proposed logic is also given for discrete-time
  Markov chains.

\end{abstract} 

\section{Introduction}

Four decades ago, Lamport~\cite{lamport77} used the notion of {\em trace 
properties} as a means to specify the correctness of individual executions of 
concurrent programs. This notion was later formalized and classified by Alpern 
and Schneider~\cite{as85} to {\em safety} and {\em liveness} properties. 
Temporal logics (e.g., \LTL~\cite{p77} and \CTL~\cite{ce81}) were built based 
on these efforts to give formal syntax and semantics to requirements of trace 
properties. Subsequently, verification algorithms were 
developed to reason about individual traces of a 
system. 

It turns out that many interesting requirements are not trace properties. For 
example, important information-flow security policies such as {\em 
noninterference}\footnote{Noninterference stipulates that input commands 
from high-privileged users have no effect on the system behavior observed by 
low-privileged observers.}~\cite{gm82} and {\em observational 
determinism}\footnote{Observational determinism requires that two executions 
that start at two low initial states appear deterministic to a low 
user.}~\cite{zm03} cannot be expressed as properties of individual execution 
traces of a system. Also, service level agreement requirements (e.g., mean 
response time and percentage uptime) that use statistics of a system across 
all executions of a system are not trace properties. Rather, they are  
properties of sets of execution traces, also known as {\em 
hyperproperties}~\cite{cs10}. Temporal logics \HyperLTL and 
\HyperCTLstar~\cite{cfkmrs14} have been proposed to provide a unifying 
framework to express and reason about hyperproperties. They allow explicit and 
simultaneous quantification over multiple paths to \LTL and to \CTLstar.

Hyperproperties can also be probabilistic. Such \emph{probabilistic 
hyperproperties} generally express probabilistic relations between independent 
executions of a system. For example, in information-flow security, adding 
probabilities is motivated by establishing a connection between information 
theory and information flow across multiple traces. It is also motivated by 
using probabilistic schedulers, which opens up an opportunity for the attacker 
to set up a probabilistic {\em covert channel}, whereby information is obtained 
by statistical inferences drawn from the relative frequency of outcomes of a 
repeated computation. Policies that defend against such an attempt, known as 
{\em probabilistic noninterference}, stipulate that the probability of every
low-observable trace be the same for every low-equivalent initial state. Such 
policies quantify on different execution traces and the probability of reaching 
certain states in the independent and simultaneous executions.

Consider the
following classic example~\cite{smith03} comprising of two threads
$\mathit{th}$ and $\mathit{th'}$:
\[
\mathit{th}:\  \mathbf{while}\ h > 0\ \mathbf{do}\ \{h \gets 
h -1\};\ l \leftarrow 2
\qquad || \qquad
\mathit{th'}:\ l \gets 1
\]
where $h$ is an input by a high-privileged user and $l$ is an output observable 
by low-privileged users. Probabilistic noninterference would require that $l$ 
obtains values of 1 and 2 with the same probability, regardless of the initial 
value of $h$. However, assuming that the scheduler chooses to execute atomic 
statements of the threads $\mathit{th}$ and $\mathit{th'}$ iteratively with 
uniform probability distribution, the likely outcome of the race between the 
two assignments $l \gets 1$ and $l \gets 2$ depends on the initial value of 
$h$: the larger the initial value of $h$, the greater the probability that the 
final value of $l$ is 2. For example, if the initial value of $h$ is 0 in one 
execution, then the final value of $l$ is 1 with probability $1/4$ and 2 with 
probability $3/4$, but for the initial value $h=5$ in another independent 
execution we can observe the final value $l=1$ with probability $1/4096$ and 
$l=2$ with probability $4095/ 4096$. Thus, it holds that for two independent 
executions with initial $h$ values $0$ resp. $5$ the larger $h$ value leads to a 
lower probability for $l=1$ upon termination. I.e., this program does not 
satisfy probabilistic noninterference.

It is straightforward to observe that requirements such as probabilistic 
noninterference cannot be expressed in existing probabilistic temporal logics 
such as \PCTL~\cite{hj94} and \PCTLstar, as they cannot draw connection 
between the probability of reaching certain states in independent executions. 
With this motivation, in this paper, we propose the temporal logic {\em 
HyperPCTL} that generalizes \PCTL by allowing explicit quantification over 
initial states and, hence, multiple computation trees simultaneously, as well 
as probability of occurring propositions that stipulate relationships among 
those traces. For the above example, the following \HyperPCTL formula expresses 
probabilistic noninterference, which obviously does not hold:
\begin{align*}\forall \sigma. \forall \sigma'. \bigg(h_\sigma \neq h_{\sigma'} 
\bigg) \; \Rightarrow \;\bigg(& \Big(\pr \F(l = 1)_\sigma = \pr (l = 
1)_{\sigma'}\Big) \, \wedge \, \\
& \Big(\pr \F(l = 2)_\sigma = \pr (l = 2)_{\sigma'}\Big)\bigg)
\end{align*}
That is, for any two executions from initial states $\sigma$ and $\sigma'$ 
(i.e., initial values of $h$), the probability distribution of terminating with 
value $1 = 1$ (or $l=2$) is uniform.
% \begin{eqnarray*}
% \forall \sigma. \forall \sigma'. &&
% \big(
% \textit{init}_\sigma \wedge \textit{init}_{\sigma'} \wedge (h=0)_\sigma \wedge (h=5)_{\sigma'} 
% \big) \; 
% \Rightarrow \;\\
% &&
% \big(
% \pr (\F(\textit{final}_{\sigma} \wedge (l = 1)_{\sigma})) > 
% \pr (\F(\textit{final}_{\sigma'}\wedge (l = 1)_{\sigma'}
% \big)\ .
% \end{eqnarray*}

In addition to probabilistic noninterference, we show that \HyperPCTL can 
express other important requirements and policies, some not related to 
information-flow security. First, we show \HyperPCTL subsumes probabilistic 
bisimulation. We also show that \HyperPCTL can express requirements such as 
differential privacy, quantitative information flow, and probabilistic causation
(a.k.a. causality). We also present a \HyperPCTL model checking algorithm for 
discrete-time Markov chains (DTMCs). The complexity of the algorithm is 
polynomial-time in the size of the input DTMC and is \comp{PSPACE-hard} in the 
size of the input \HyperPCTL formula. We also discuss a wide range of open 
problem to be tackled by future research. We believe that this paper opens a 
new area in rigorous analysis of probabilistic systems.

\paragraph{Organization} The rest of the paper is organized as follows. 
Section~\ref{sec:hpctl} defines the syntax and semantics of \HyperPCTL. 
Section~\ref{sec:examples} provides a diverse set of example requirements that 
\HyperPCTL can express. We present our model checking algorithm in 
Section~\ref{sec:mc}. Related work is discussed in Section~\ref{sec:related}. 
Finally, we make concluding remarks and discuss future work in 
Section~\ref{sec:concl}. 
\section{HyperPCTL}
\label{sec:hpctl}

In this section, we present the syntax and semantics of \HyperPCTL.

\begin{definition}
A {\em (discrete-time) Markov chain (DTMC)} $\dtmc = (S, \tpm, \AP, 
L)$ is a tuple with the following components:

\begin{itemize}
\item $S$ is a finite nonempty set of {\em states},

\item $\tpm : S \times S \rightarrow [0, 1]$ is a {\em transition probability 
function} with $\sum_{s' \in S} \tpm(s, s') =1$ for all states $s \in S$,

\item $\AP$ is a set of {\em atomic propositions}, and 

\item $L : S \rightarrow 2^{\AP}$ is a \emph{labeling function}.\hfill\qed
\end{itemize}

\end{definition}

%The value $\init$ specifies the probability that the system evaluation starts 
%in a state $s$. The states $s$ with $\init(s) > 0$ are considered as possible 
%{\em initial states}.

A \emph{path} of a Markov chain $\dtmc = (S, \tpm, \AP, L)$ is defined as an 
infinite sequence $\pi = s_0s_1s_2\cdots \in S^\omega$ of states with
$\tpm(s_i, s_{i+1}) > 0$, for all $i \geq 0$; we write $\pi[i]$ for $s_i$. Let 
$\Paths{s}{\dtmc}$ denote the set of all (infinite) paths starting in $s$ in 
$\dtmc$, and $\fPaths{s}{\dtmc}$ denote the set of all finite prefixes of paths 
from $\Paths{s}{\dtmc}$, which we sometimes call \emph{finite paths}.

\subsection{Syntax}

\HyperPCTL \emph{state formulas} are inductively defined by the following 
grammar:
\[
\begin{array}{lll}
\psi & ::= & \forall \sigma.\psi~~\Big\vert~~\exists 
\sigma.\psi ~~\Big\vert~~ \tru ~~\Big\vert~~ a_\sigma~~\Big\vert~~\psi \wedge 
\psi~~\Big\vert~~\neg 
\psi~~\Big\vert~~p\sim p \\
p & ::= & \pr(\varphi) ~~\Big\vert~~ c ~~\Big\vert~~ p+p ~~\Big\vert~~ p-p ~~\Big\vert~~ p\cdot p
\end{array}
\]
where $c\in\mathbb{Q}$, $a \in \AP$ is an atomic proposition, $\sim \in \{<, 
\leq, =, \geq, >\}$, $\sigma$ is a {\em state variable} from a countably 
infinite supply of variables $\mathcal{V}=\{\sigma_1,\sigma_2,\ldots\}$, $p$ is 
a \emph{probability expressions}, and $\varphi$ is a {\em path formula}. 
\HyperPCTL path formulas are formed according to the following grammar:
$$\varphi~~ ::=~~ \Next \psi ~~\Big\vert~~ \psi \, \U \, \psi
~~\Big\vert~~ \psi \, \U^{[k_1,k_2]} \,\psi$$
where $\psi$ is a state formulas and $k_1, k_2 \in \naturalszero$ with $k_1\leq 
k_2$. 

We also introduce state formulas of the form $p\in J$, where $J = [l,u] 
\subseteq [0, 1]$ is an interval with rational bounds, as syntactic sugar for 
$l\leq p \wedge p \leq u$. We also define the syntactic sugar $ \psi_1 \, 
\U^{\leq k} \,\psi_2$ for $\psi \, \U^{[0,k]} \,\psi$. Also, $\psi_1 \vee 
\psi_2 = \neg(\neg \psi_1 \wedge \neg\psi_2)$, $\F \psi = \tru \, \U \, \psi$, 
$\G \psi = \neg \F \neg \psi$, $\F^{[k_1,k_2]} \psi = \tru \, \U^{[k_1,k_2]} \, 
\psi$, and $\G^{[k_1,k_2]} \psi = \neg \F^{[k_1,k_2]} \neg \psi$. We denote by 
$\mathcal{F}$ the set of all \HyperPCTL state formulas.

An occurrence of an indexed atomic proposition $a_ \sigma$ in a \HyperPCTL state 
formula $\psi$ is \emph{free} if it is not in the scope of a quantifier bounding 
$\sigma$ and otherwise \emph{bound}; we denote by $\textit{Free}(\psi)$ the set 
of all indexed atomic propositions with at least one free occurrence in $\psi$. 
\HyperPCTL \emph{sentences} are \HyperPCTL state formulas in which all 
occurrences of all indexed atomic propositions are bound. \HyperPCTL 
(\emph{quantified}) \emph{formulas} are \HyperPCTL sentences. 

\paragraph{Example} Consider the formula
$$\forall \sigma_1.\exists \sigma_2. \pr(\F a_{\sigma_1})=\pr(\F 
b_{\sigma_2}).$$
This formula is true if for each state $s_1$ there
exists another state $s_2$ such that the probability to finally reach a state labelled with $a$ from $s_1$ equals the probability of reaching $b$ from $s_2$.

\subsection{Semantics}

We present the semantics of \HyperPCTL based on $n$-ary self-composition of a 
DTMC. We emphasize that it is possible to define the semantics in terms of the 
non-self-composed DTMC, but it will essentially result in a very similar 
setting, but more difficult to understand.

\begin{definition}
The \emph{n-ary self-composition} of a DTMC $\dtmc = (S, \tpm, \AP, L)$ is a 
DTMC $\dtmc^n = (S^n, \tpm^n, \AP^n, L^n)$ with

\begin{itemize}

\item $S^n=S\times\ldots\times S$ is the $n$-ary Cartesian product of $S$,

\item $\tpm^n\big(s, s')=\tpm(s_1,s_1'\big)\cdot\ldots\cdot \tpm(s_n,s_n')$ for 
all $s=(s_1,\ldots,s_n)\in S^n$ and $s'=(s_1',\ldots,s_n')\in S^n$,

\item $\AP^n=\cup_{i=1}^n\AP_i$, where $\AP_i=\{a_i\,|\,a\in\AP\}$ for $i 
\in [1, n]$, and

\item $L^n(s) = \cup_{i=1}^n L_i(s_i)$ for all $s = (s_1,\ldots,s_n)\in S^n$ 
with $L_i(s_i) = \{a_i\,|\,a\in L(s_i)\}$ for $i \in [1, n]$.\hfill\qed
\end{itemize}

\end{definition}

\noindent The satisfaction of a \HyperPCTL quantified formula by a DTMC $\dtmc = (S, \tpm, \AP, L)$ is defined by:
\[
\dtmc \models \psi \qquad 
\textit{iff} \qquad
\dtmc,()\models\psi
\]
\noindent where $()$ is the empty sequence of states.  Thus, the satisfaction 
relation $\models$ defines the values of \HyperPCTL quantified, state, and path 
formulas in the context of a DTMC $\dtmc = (S, \tpm, \AP, L)$ and an $n$-tuple 
$s=(s_1,\ldots,s_n)\in S^n$ of states (which is $()$ for $n=0$). Intuitively, 
the state sequence $s$ stores instantiations for quantified state variables.  
Remember that \HyperPCTL quantified formulas are sentences.  The semantics 
evaluates \HyperPCTL formulas by structural recursion. Quantifiers are 
instantiated and the instantiated values for state variables are stored in the 
state sequence $s$. To maintain the connection between a state in this sequence 
and the state variable which it instantiates, we introduce the auxiliary syntax 
$a_i$ with $a\in\AP$ and $i \in \naturals$, and if we instantiate $\sigma$ in
$\exists\sigma.\psi$ or $\forall\sigma.\psi$ by state $s$, then we append $s$ 
at the end of the state sequence and replace all $a_\sigma$ that is bound by 
the given quantifier by $a_i$ with $i$ being the index of $s$ in the state 
sequence. We will express the meaning of path formulas based on the $n$-ary 
self-composition of $\dtmc$; the index $i$ for the instantiation of $\sigma$ 
also fixes the component index in which we keep track of the paths starting in 
$\sigma$. The semantics judgment rules are the following:
\[
\begin{array}{l@{\quad}c@{\quad}l}
\dtmc,s \models \forall \sigma.\psi & 
\textit{iff} & 
\forall s_{n+1} \in S.\ \dtmc,(s_1,\ldots,s_n, s_{n+1}) \models \psi[\AP_{n+1}/\AP_{\sigma}]\\
\dtmc,s \models \exists \sigma.\psi & 
\textit{iff} & 
\exists s_{n+1} \in S.\ \dtmc,(s_1,\ldots,s_n, s_{n+1}) \models \psi[\AP_{n+1}/\AP_{\sigma}]\\
\dtmc,s \models \texttt{true} &
 &
\\
\dtmc,s \models a_i &
\textit{iff} &
a \in L(s_i)\\
\dtmc,s  \models \psi_1 \wedge \psi_2 &
\textit{iff} &
\dtmc,s \models \psi_1 \textit{ and } \dtmc,s  \models \psi_2\\
\dtmc,s  \models \neg \psi &
\textit{iff} &
\dtmc,s \not \models \psi\\
\dtmc,s \models p_1 \sim p_2 &
\textit{iff} &
\llbracket p_1\rrbracket_{\dtmc,s} \sim \llbracket p_2 \rrbracket_{\dtmc,s}\\
\llbracket \pr(\varphi)\rrbracket_{\dtmc,s} & = & \Pr\{\pi \in \Paths{s}{\dtmc^n} \mid \dtmc,\pi \models \varphi\}\\
\llbracket c\rrbracket_{\dtmc,s} & = & c\\
\llbracket p_1+p_2\rrbracket_{\dtmc,s} & = & \llbracket p_1\rrbracket_{\dtmc,s} + \llbracket p_2\rrbracket_{\dtmc,s}\\
\llbracket p_1-p_2\rrbracket_{\dtmc,s} & = & \llbracket p_1\rrbracket_{\dtmc,s} - \llbracket p_2\rrbracket_{\dtmc,s}\\
\llbracket p_1\cdot p_2\rrbracket_{\dtmc,s} & = & \llbracket p_1\rrbracket_{\dtmc,s} \cdot \llbracket p_2\rrbracket_{\dtmc,s}
\end{array}
\]

\noindent where $\psi$, $\psi_1$ and $\psi_2$ are \HyperPCTL state formulas; 
the substitution $\psi[\AP_{n+1}/\AP_{\sigma}]$ replaces for each atomic
proposition $a\in\AP$ each free occurrence of $a_{\sigma}$ in $\psi$ by
$a_{n+1}$; $a\in\AP$ is an atomic proposition and $1\leq i \leq n$; $p_1$ and 
$p_2$ are probability expressions and $\sim \in \{<, \leq, =, \geq, >\}$; 
$\varphi$ is a \HyperPCTL path formula and $c$ is a rational constant.

The satisfaction relation for \HyperPCTL path formulas is defined as follows, 
where $\pi$ is a path of $\dtmc^n$ for some $n\in\naturals$; $\psi$, $\psi_1$, 
and $\psi_2$ are \HyperPCTL state formulas and $k\in\naturalszero$:
\[
\begin{array}{l@{\quad}c@{\quad}l}
\dtmc,\pi \models \X \psi &
\textit{iff} &
\dtmc,\pi[1] \models \psi\\
\dtmc,\pi \models \psi_1 \U \psi_2 &
\textit{iff} &
\exists j \geq 0. \Big(\dtmc,\pi[j]\models\psi_2 \wedge \forall i \in [0, j). \dtmc,\pi[i] \models \psi_1\Big)\\
\dtmc,\pi \models \psi_1 \U^{[k_1,k_2]} \, \psi_2 &
\textit{iff} &
\exists j \in [k_1, k_2]. \Big(\dtmc,\pi[j]\models\psi_2 \wedge \forall i \in [0, j). \dtmc,\pi[i] \models \psi_1\Big)
\end{array}
\]
Note that the semantics assures that each path formula $\varphi$ is evaluated in the context of a path of $\dtmc^n$ such that $1\leq i \leq n$ for each $a_i$ in $\varphi$.

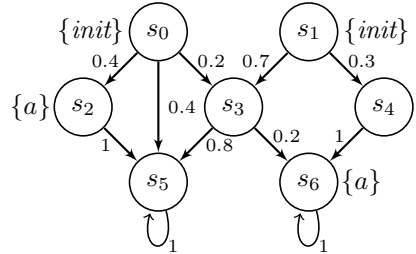
\begin{wrapfigure}{r}{6cm}
\centering
\begin{tikzpicture}

\node[draw,circle,text width=0.5cm] (s0) at (0,0) {};
\node at (0,0) {$s_0$};
\node[draw,circle,text width=0.5cm] (s1) at (2,0) {};
\node at (2,0) {$s_1$};
\node[draw,circle,text width=0.5cm] (s2) at (-1,-1) {};
\node at (-1,-1) {$s_2$};
\node[draw,circle,text width=0.5cm] (s3) at (1,-1) {};
\node at (1,-1) {$s_3$};
\node[draw,circle,text width=0.5cm] (s4) at (3,-1) {};
\node at (3,-1) {$s_4$};
\node[draw,circle,text width=0.5cm] (s5) at (0,-2) {};
\node at (0,-2) {$s_5$};
\node[draw,circle,text width=0.5cm] (s6) at (2,-2) {};
\node at (2,-2) {$s_6$};

\node[left of=s0,node distance=0.9cm] {$\{\textit{init}\}$};
\node[right of=s1,node distance=0.9cm] {$\{\textit{init}\}$};
\node[left of=s2,node distance=0.7cm] {$\{a\}$};
\node[right of=s6,node distance=0.7cm] {$\{a\}$};

\path[draw,-latex',thick] (s0) edge node[left,font=\scriptsize,near start] {$0.4$} (s2);
\path[draw,-latex',thick] (s0) edge node[right,font=\scriptsize,near start=5pt] {$0.2$} (s3);
\path[draw,-latex',thick] (s0) edge node[right,font=\scriptsize] {$0.4$} (s5);
\path[draw,-latex',thick] (s1) edge node[left,font=\scriptsize,near start] {$0.7$} (s3);
\path[draw,-latex',thick] (s1) edge node[right,font=\scriptsize,near start] {$0.3$} (s4);
\path[draw,-latex',thick] (s2) edge node[left,font=\scriptsize] {$1$} (s5);
\path[draw,-latex',thick] (s3) edge node[right,font=\scriptsize] {$0.8$} (s5);
\path[draw,-latex',thick] (s3) edge node[right,font=\scriptsize,near start] {$0.2$} (s6);
\path[draw,-latex',thick] (s4) edge node[left,font=\scriptsize,near start] {$1$} (s6);
\path[draw,-latex'] (s5) edge [loop below] node[right,font=\scriptsize] {$1$} (s5);
\path[draw,-latex'] (s6) edge [loop below] node[right,font=\scriptsize] {$1$} (s6);
\end{tikzpicture}
\vspace*{-2ex}
 \caption{Semantics example.}
 \label{fig:semantics}
\vspace*{-5ex}
\end{wrapfigure}

\paragraph{Example} Consider the DTMC $\dtmc$ in Fig.~\ref{fig:semantics} and 
the following \HyperPCTL formula:
$$
\psi = \forall \sigma.\forall \sigma'.  (\textit{init}_\sigma\wedge\textit{init}_{\sigma'}) \Rightarrow
  \Big(\pr(\F a_{\sigma}) = \pr(\F 
a_{\sigma'})\Big)
$$
\noindent This formula is satisfied by $\dtmc$ if for all pairs of initial states (labelled by the atomic proposition $\textit{init}$) the probability to satisfy $a$ is the same, i.e., for each $(s_i,s_j)\in S^2$ with $\textit{init}\in L(s_i)$ and $\textit{init}\in L(s_j)$ it holds that $\dtmc,(s_i,s_j)\models\pr(\F a_{1}) = \pr(\F 
a_{2}) $.  The probability of reaching $a$ from $s_0$ is $0.4 + (0.2 \times 
0.2) = 0.44$. Moreover, the probability of reaching $a$ from $s_1$ is $0.3 + 
(0.7 \times 0.2) = 0.44$. Hence, we have $\dtmc \models \psi$.

\section{HyperPCTL in Action}
\label{sec:examples}

We now put \HyperPCTL into action by formulating probabilistic requirements 
from different areas, such as information-flow security, privacy, and 
causality analysis.

\subsection{Probabilistic Bisimulation}
\label{subsec:bisim}

A \emph{bisimulation} is an equivalence relation over a set of states of a 
system such that equivalent states cannot be distinguished by observing their 
behaviors. In the context of DTMC states and PCTL properties, a
\emph{probabilistic} bisimulation is an equivalence relation over the DTMC
states such that any two equivalent states satisfy the same PCTL formulas. The 
latter property can be assured inductively by requiring that equivalent states 
have the same labels and the probability to move from them to any of the 
equivalence classes is the same.

Assume a partitioning $S_1,\ldots,S_k$ of $S$ with $\cup_{i=1}^k S_i = S$ and 
$S_i \cap S_j=\emptyset$ for all $1\leq i<j\leq k$.  To express that the 
equivalence relation $R=\cup_{i=1}^k S_i\times S_i$ is a bisimulation, we define
$\dtmc'=(S,\tpm,\AP',L')$ with $\AP'=\AP \, \cup \, \{a^1,\ldots,a^k\}$, where 
each $a^i$, for all $i \in [1, k]$, is a fresh atomic proposition not in $\AP$, 
and for each $s \in S_i$, we set $L'(s) = L(s) \cup \{a^i\}$. The equivalence 
relation $R$ is a bisimulation for $\dtmc$ if $\dtmc'$ satisfies the following 
\HyperPCTL formula
  \[
\varphi_{\mathsf{pb}} = \forall\sigma.\forall\sigma'. \bigwedge_{i=1}^{k}\left[
    (a^i_\sigma  \wedge a^i_{\sigma'})  \Rightarrow \Bigg[ \psi^{\AP}  \wedge 
 \pr\Bigg( 
  \G \bigwedge_{j=1}^{k} 
  \bigg( 
    \pr(
      \X 
a^j_{\sigma}
    )
    =
    \pr(
      \X 
      a^j_{\sigma'}
    )
  \bigg)
\Bigg) = 1\Bigg]\right]
\]
where $\psi^{\AP}=\bigwedge_{a\in\AP} (a_{\sigma}\leftrightarrow a_{\sigma'})$.

\subsection{Probabilistic Noninterfence}

{\em Noninterference} is an information-flow security policy that enforces that 
a low-privileged user (e.g., an attacker) should not be able to distinguish two 
computations from their publicly observable outputs if they only vary in their 
inputs by a high-privileged user (e.g., a secret). {\em Probabilistic 
noninterference}~\cite{g92} establishes connection between information theory 
and information flow by employing probabilities to address covert channels.  
Intuitively, it requires that the probability of every 
low-observable trace pattern is the same for every low-equivalent initial state. 
Probabilistic noninterference can be expressed in \HyperPCTL as follows:
$$\varphi_{\mathsf{pni}}= \forall \sigma. \forall \sigma'. \Big(l_\sigma \wedge 
l_{\sigma'}\Big) \Rightarrow \pr\bigg[\G\Big(\pr(\X l_\sigma) = \pr(\X 
l_{\sigma'})\Big) \bigg] = 1$$
where $l$ denotes a low-observable atomic proposition. Observe that formula 
$\varphi_{\mathsf{pni}}$ is a simplification of formula $\varphi_{\mathsf{pb}}$ 
in Section~\ref{subsec:bisim}. In fact, most approaches to prove probabilistic 
noninterference is by showing probabilistic bisimulation with respect to 
low-observable propositions.

\subsection{Quantitative Information Flow}

Roughly speaking, the quantitative information flow (QIF) problem is concerned 
with the amount of information that an attacker can learn about the high 
security input by executing the program and observing the low security output, 
i.e., the difficulty of guessing the secret input from the channel output. 
QIF aims at quantifying the amount of information in a high security input $H$ 
(e.g., a password), which is the attacker's initial uncertainty, from the 
amount of information leaked to a low security output $L$ (e.g., the result of 
a password verification), and the amount of unleaked information about $H$ (the 
attacker's remaining uncertainty). For example, consider the following two 
programs:
$$P_1 \; \triangleq \; \mathbf{if}~H = c~~{\bf then}~L 
:= 1~~\mathbf{else}~L := 0$$
and
$$P_2 \; \triangleq \; L := H$$
Although discrete noninterference would characterize both as insecure, QIF 
characterizes $P_1$ as ``more'' secure than $P_2$.

QIF is founded on information-theoretic concepts (e.g., 
Shannon/min/guess-entropy) that can compute the amount of 
leaked bits of a high security input~\cite{chm05,cms09,g92,kb07}. The {\em 
bounding problem}~\cite{yt11} is to determine whether that amount is 
bounded from above by a constant $q$. Assuming that the values of the high 
input security $H$ to the program are uniformly distributed, each given in a 
different initial state, and the program outputs a value from $L$ only once at 
termination, a simple QIF policy~\cite{dkmr15} is (1) starting from any high 
input, the probability of reaching any of the low-observable output is 
bounded by $\frac{\log|L|}{\log |H|}$, and (2) starting from every pair of 
high inputs, the probability of reaching the same low-observable output is the 
same. This QIF policy can be formulated in \HyperPCTL as follows:
$$\psi_\mathsf{qif} = \forall \sigma.\forall\sigma'. \bigg(\bigwedge_{l \in 
L}\pr(\F l_\sigma) \leq \frac{\log |L|}{\log |H|}\bigg) \; \wedge \; 
\bigg(\bigwedge_{l \in L}\pr(\F 
l_\sigma) = \pr(\F l_{\sigma'})\bigg)$$
where $l$ labels states where a different low security output from $L$ is 
observed.

\subsection{Differential Privacy}

{\em Differential privacy}~\cite{dr14} is a commitment by a data holder to a 
data subject (normally an individual) that he/she will not be affected by 
allowing his/her data to be used in any study or analysis. Formally, let 
$\epsilon$ be a positive real number and $\mathcal{A}$ be a randomized algorithm 
that makes a query to an input database and produces an output. Algorithm 
$\mathcal{A}$ is called {\em $\epsilon$-differentially private}, if for all 
databases $D_1$ and $D_2$ that differ on a single element, and all subsets $S$ 
of possible outputs of $\mathcal{A}$, we have:
$$\Pr[\mathcal{A}(D_{1})\in S] \leq e^{\epsilon } \cdot 
\Pr[\mathcal{A}(D_{2})\in S].$$

Differential privacy can be expressed in \HyperPCTL by the following formula:
\begin{align*}
\label{eq:dp}
\nonumber \psi_{\mathsf{dp}} = \forall \sigma. \forall \sigma'. & 
\bigg[\dbsim(\sigma, \sigma')\bigg] \; \Rightarrow \\
& \bigg[\pr\Big(\F (\qout \in S)_\sigma\Big) \leq e^\epsilon \cdot \pr\Big(\F 
(\qout \in S)_{\sigma'}\Big)\bigg]
\end{align*}
where $\dbsim(\sigma, \sigma')$ means that two different dataset inputs have 
all but one similarity and $\qout$ is the result of the query. For example, one 
way to provide differential privacy is through {\em randomized 
response} in order to create noise and provide plausible deniability. Let $A$ 
be an embarrassing or illegal activity. In a social study, each participant is 
faced with the query, ``Have you engaged in activity $A$ in the past week?'' and 
is instructed to respond by the following protocol:

\begin{enumerate}
\item Flip and coin.
\item If tail, then answer truthfully.
\item If head, then flip the coin again and respond ``Yes'' if head and ``No''
if tail.
\end{enumerate}
Thus, a ``Yes'' response may have been offered because the first and second 
coin flips were both heads. This implies that, there are no good or bad 
responses and an answer cannot be incriminating.

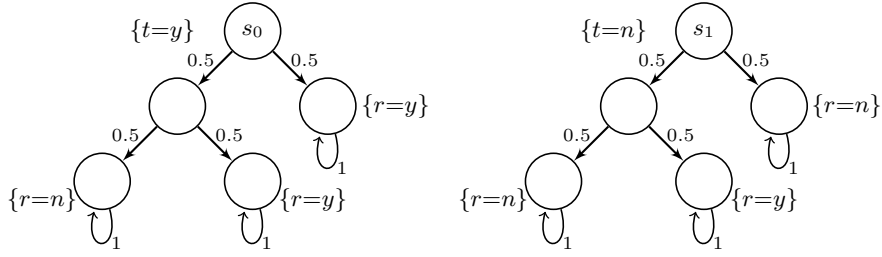
\begin{figure}[t]
\centering
\begin{tikzpicture}

\begin{scope}
\node[draw,circle,text width=0.5cm] (s0) at (0,0) {};
\node at (0,0) {$s_0$};
\node[draw,circle,text width=0.5cm] (s1) at (-1,-1) {};
\node[draw,circle,text width=0.5cm] (s2) at (1,-1) {};
\node[draw,circle,text width=0.5cm] (s3) at (-2,-2) {};
\node[draw,circle,text width=0.5cm] (s4) at (0,-2) {};

\node[left of=s0,node distance=1.2cm] {$\{t{=}y\}$};
\node[right of=s2,node distance=0.9cm] {$\{r{=}y\}$};
\node[left of=s3,node distance=0.8cm,yshift=-7pt] {$\{r{=}n\}$};
\node[right of=s4,node distance=0.8cm,yshift=-7pt] {$\{r{=}y\}$};

\path[draw,-latex',thick] (s0) edge node[left,font=\scriptsize,near start] {$0.5$} (s1);
\path[draw,-latex',thick] (s0) edge node[right,font=\scriptsize,near start] {$0.5$} (s2);
\path[draw,-latex',thick] (s1) edge node[left,font=\scriptsize,near start] {$0.5$} (s3);
\path[draw,-latex',thick] (s1) edge node[right,font=\scriptsize,near start] {$0.5$} (s4);
\path[draw,-latex'] (s2) edge [loop below] node[right,font=\scriptsize] {$1$} (s2);
\path[draw,-latex'] (s3) edge [loop below] node[right,font=\scriptsize] {$1$} (s3);
\path[draw,-latex'] (s4) edge [loop below] node[right,font=\scriptsize] {$1$} (s4);
\end{scope}

\begin{scope}[xshift=6cm]
\node[draw,circle,text width=0.5cm] (s0) at (0,0) {};
\node at (0,0) {$s_1$};
\node[draw,circle,text width=0.5cm] (s1) at (-1,-1) {};
\node[draw,circle,text width=0.5cm] (s2) at (1,-1) {};
\node[draw,circle,text width=0.5cm] (s3) at (-2,-2) {};
\node[draw,circle,text width=0.5cm] (s4) at (0,-2) {};

\node[left of=s0,node distance=1.2cm] {$\{t{=}n\}$};
\node[right of=s2,node distance=0.9cm] {$\{r{=}n\}$};
\node[left of=s3,node distance=0.8cm,yshift=-7pt] {$\{r{=}n\}$};
\node[right of=s4,node distance=0.8cm,yshift=-7pt] {$\{r{=}y\}$};

\path[draw,-latex',thick] (s0) edge node[left,font=\scriptsize,near start] {$0.5$} (s1);
\path[draw,-latex',thick] (s0) edge node[right,font=\scriptsize,near start] {$0.5$} (s2);
\path[draw,-latex',thick] (s1) edge node[left,font=\scriptsize,near start] {$0.5$} (s3);
\path[draw,-latex',thick] (s1) edge node[right,font=\scriptsize,near start] {$0.5$} (s4);
\path[draw,-latex'] (s2) edge [loop below] node[right,font=\scriptsize] {$1$} (s2);
\path[draw,-latex'] (s3) edge [loop below] node[right,font=\scriptsize] {$1$} (s3);
\path[draw,-latex'] (s4) edge [loop below] node[right,font=\scriptsize] {$1$} (s4);
\end{scope}

\end{tikzpicture}
\caption{Markov chain of the randomized response protocol.}
\label{fig:yesno}
\end{figure}

We now show that this social study is $(\ln 3)$-deferentially private. For 
each participant in the study, Fig.~\ref{fig:yesno} shows the Markov chain of 
the response protocol, where $\{t{=}y\}$ (respectively, $\{t{=}n\}$) denotes that 
the truth is that the participant did (respectively, did not) engage in 
activity $A$, and $\{r{=}y\}$ (respectively, $\{r{=}n\}$) means that the participant 
responds ``Yes'' (respectively, ``No''). The \HyperPCTL formula is the 
following:
\begin{align*}
\forall \sigma. \forall \sigma'. & \bigg[\bigg((t {=} n)_\sigma \, \wedge \,
(t {=} y)_{\sigma'}\bigg) \; \Rightarrow \; \bigg(\pr \Big(\F (r {=} 
n)_\sigma\Big) \leq  e^{\ln 3} \cdot \pr\Big(\F (r {=} 
n)_{\sigma'}\Big)\bigg)\bigg] \; \wedge \\
& \bigg[\bigg((t {=} y)_\sigma \, \wedge \, (t {=} n)_{\sigma'}\bigg) \; 
\Rightarrow \; \bigg(\pr\Big(\F (r {=} y)_\sigma\Big)  \leq e^{\ln 3} \cdot 
\pr\Big(\F (r {=} y)_{\sigma'}\Big)\bigg)\bigg]
\end{align*}

Observe that compared to formula $\psi_\mathsf{dp}$, we have 
decomposed $\dbsim(\sigma, \sigma')$ to two cases of $t = y$ and $t = n$. Thus, 
in the left conjunct, the set $S$ represents the case where the response is 
``Yes'' and in the right conjunct, the set $S$ represents the case where the 
response is ``No''. It is straightforward to see that the DTMC in 
Fig.~\ref{fig:yesno} satisfies the formula, when for the left conjunct $\sigma$ 
and $\sigma'$ are instantiated by $s_0$ and $s_1$, respectively, and for the 
right conjunct $\sigma$ and $\sigma'$ are instantiated by $s_1$ and $s_0$, 
respectively. 

%https://www.cis.upenn.edu/~aaroth/Papers/privacybook.pdf

\subsection{Probabilistic Causation}

{\em Probabilistic causation}~\cite{f88} aims to characterize the relationship 
between {\em cause} and {\em effect} using the tools of probability theory. The 
reason for using probabilities is that most causes are not invariably followed 
by their effects. For example, smoking is a cause of lung cancer, even though 
some smokers do not develop lung cancer and some people who have lung cancer are 
not smokers. Thus, we need to somehow express that some causes are {\em more 
likely} to develop an effect. Specifically, the central idea in probabilistic 
causation is to assert that the probability of occurring effect $e$ if cause $c$ 
happens is higher than the probability of occurring $e$ when $c$ does not 
happen. We can express the most basic type of probabilistic causation in 
\HyperPCTL as follows:
$$\psi_{\mathsf{pc_1}} = \forall \sigma.\forall \sigma'. \bigg(\pr(c_\sigma 
\, \, \U e_\sigma) > \pr(\neg c_{\sigma'} \, \U e_{\sigma'})\bigg).$$
Observe that we assume that the occurrence or absence of the cause is 
persistent (hence, the until operator). Also, expressing causation in the 
standard \PCTL by stripping the state quantifiers in formula 
$\psi_{\mathsf{pc_1}}$ will damage the meaning of causation. The resulting 
\PCTL formula captures the causation relation from each initial state in 
isolation and it wrongly allows the probability of $(c \, \U \, e)$ from one 
initial state to be less than the probability of $(\neg c \, \U \, e)$ from 
another initial state. 

One problem with formula $\psi_{\mathsf{pc_1}}$ is spurious correlations. 
For example, if $c$ is the drop in the level of mercury in a barometer, and $e$ 
is the occurrence of a storm, then the above formula may hold in a system, 
though $c$ is not really the cause of $e$. In fact, the real cause for both is 
the drop in atmospheric pressure. To address this problem, we add a constraint, 
where there should be no further event $a$ that {\em screens off} $e$ from 
$c$~\cite{r56}:
\begin{align*}
\psi_{\mathsf{pc_2}} = \forall \sigma.\forall \sigma'. \neg \exists 
\sigma''. & \bigg(\pr(c_\sigma \, \U \, e_\sigma) > \pr(\neg c_{\sigma'} 
\, \U e_{\sigma'})\bigg) \, \wedge \\
\bigwedge_{a \in \AP\setminus\{e,c\}} & \bigg(\pr((a_{\sigma''} \wedge 
c_{\sigma''}) \, \U \, e_{\sigma''}) = \pr(c_\sigma \, \U \, 
e_\sigma)\bigg).
\end{align*}
%The negation behind the existential quantifier can be pushed inside to obtain a 
%proper \HyperPCTL formula. 
The negation behind the existential quantifier can be pushed inside to obtain 
a proper \HyperPCTL formula. We note that for simplicity, in formula 
$\psi_{\mathsf{pc_2}}$, propositions $a$ and $c$ occur in the same state in 
$\sigma''$. A more general way is to allow $a$ happen before or simultaneously 
with $c$. Finally, we note that other concepts in probabilistic causation such 
as Reichenbach's Common Cause Principle and Fork Asymmetry~\cite{r56} (which 
emulates the second law of thermodynamics), as well as Skyrms's Background 
Contexts~\cite{s80} can be expressed in a similar fashion.

\section{HyperPCTL Model Checking}
\label{sec:mc}

In the following, we show that the \HyperPCTL model checking problem is 
decidable by introducing a model checking algorithm. The space complexity of 
our algorithm is exponential in the number of quantifiers of the input formula, 
because for $n$ state quantifiers, we build the $n$-ary self-composition of the 
input DTMC. We are uncertain whether there exists a \comp{PSPACE} algorithm, 
but we show the \comp{PSPACE-hardness} of the problem.

Let $\dtmc= (S, \tpm, \AP, L)$ be a DTMC and $\psi$ be a \HyperPCTL quantified 
formula with $n$ state quantifiers. Informally, our model checking algorithm 
decides whether $\dtmc\models\psi$ as follows (detailed pseudo-code is 
formulated in the Algorithms \ref{algo:mc1}--\ref{algo:mc3}):

\begin{enumerate}

\item If $n=0$, then $\psi$ contains constants only; evaluate $\psi$ and return 
the result.

 \item Otherwise, if $n>0$, then apply variable renaming such that the 
quantified state variables are named $\sigma_1,\ldots,\sigma_n$.

\item Build the self-composition $\dtmc^n$.

 \item Compute a labeling $\hat{L}^n(s)$ for all states $s\in S^n$ of $\dtmc^n$ 
as follows. Initially $\hat{L}^n(s)=\emptyset$ for all $s\in S^n$ 
(Line~\ref{line:empty} in Algorithm~\ref{algo:mc1}). For all sub-formulas 
$\psi'$ of $\psi$ inside-out do the following:

\begin{itemize}

\item If the subformula $\psi'$ has the form $\tru$, add $\tru$ to the label 
sets $\hat{L}^n(s)$ of all states $s\in S^n$ (Line~\ref{line:true} in 
Algorithm~\ref{algo:mc2}).

\item If the subformula $\psi'$ is an atomic proposition $a_{\sigma_i}$, add 
$a_{\sigma_i}$ to the label set of each state $s\in S^n$ with $a_i\in L^n(s)$ 
(Line~\ref{line:prop} in Algorithm~\ref{algo:mc2}).

\item If the subformula $\psi'$  is $\psi_1 \, \wedge \, \psi_2$, then add 
$\psi_1 \, \wedge \, \psi_2$ to $\hat{L}^n(s)$ for each $s\in S^n$ with 
$\psi_1\in \hat{L}^n(s)$ and $\psi_2\in \hat{L}^n(s)$ 
(Lines~\ref{line:and1} -- \ref{line:and2} in Algorithm~\ref{algo:mc2}).

\item If the subformula $\psi'$ is $\neg\psi_1$, then add $\neg\psi_1$ to 
$\hat{L}^n(s)$ for each $s\in S^n$ with $\psi_1\not\in \hat{L}^n(s)$ 
(Lines~\ref{line:neg1} -- \ref{line:neg2} in Algorithm~\ref{algo:mc2}).

\item If the subformula $\psi'$ is $p_1\sim p_2$ (respectively $p\in J$), then
compute for all $\pr(\varphi)$ appearing in $p_1\sim p_2$ (respectively, $p\in 
J$) for all states $s\in S^n$ the probability that $\varphi$ holds in $s$ using 
standard \PCTL model checking, and add for all $s\in S^n$ the property $p_1\sim 
p_2$ (respectively, $p\in J$) to $\hat{L}^n(s)$ if $p_1\sim p_2$ (respectively, 
$p\in J$) evaluates to $\tru$ in $s$ (Lines~\ref{line:pctl1} -- 
\ref{line:pctl2} in Algorithm~\ref{algo:mc2}).

\item If the subformula $\psi'$ is of the form $\exists \sigma_i.\psi_1$, then 
label all states $s = (s_1,\ldots,s_n) \in S^n$ with $\exists \sigma_i.\psi_1$ 
iff there exists an $s_i'\in S$, such that $\psi_1 \in \hat{L}^n(s_1, \ldots, 
s_{i-1},s_i',s_{i+1},\ldots,s_n)$ (Lines~\ref{line:exists1} -- 
\ref{line:exists2} in Algorithm~\ref{algo:mc2}).

\item If the subformula $\psi'$ is of the form $\forall \sigma_i.\psi_1$, then 
label all states $s = (s_1,\ldots,s_n)\in S^n$ with $\forall\sigma_i.\psi_1$ 
iff for all $s_i'\in S$ it holds that $\psi_1 \in \hat{L}^n(s_1, \ldots, 
s_{i-1}, s_i', s_{i+1}, \ldots, s_n)$ (Lines~\ref{line:forall1} -- 
\ref{line:forall2} in Algorithm~\ref{algo:mc2}).
\end{itemize}

\item Upon termination of the above iterative labeling procedure, as
  $\psi$ is a sentence and thus state-independent, either all states
  are labelled with it or none of them. Return $\tru$ if for an
  arbitrary state $s$ we have $\psi\in\hat{L}^n(s)$ and return $\fals$ otherwise..
\end{enumerate}
\vspace*{-1ex}

\begin{algorithm}[t]
  \caption{\HyperPCTL model checking algorithm I}
  \label{algo:mc1}
  \SetKwFunction{main}{main}
  \Input{DTMC $\dtmc=(S, \tpm, \AP, L)$, \HyperPCTL quantified formula $\psi$}
  \Output{Whether $\dtmc\models\psi$}
  \Fn{main{$(\dtmc,\psi)$}}{
    $n$ := number of quantifiers in $\psi$\;
    let $\hat{L}^n:S^n\rightarrow 2^\mathcal{F}$ with $\hat{L}^n(s)=\emptyset$ 
for all $s\in S^n$ \label{line:empty}\;
    $\hat{L}^n(s)$ := HyperPCTL($\dtmc,\psi,n,\hat{L}^n$) \hfill {\em \% see 
Algorithm~\ref{algo:mc2}}\;
    \If{$\psi\in\hat{L}^n(s)$ for some $s\in S^n$}{\Return{$\tru$}}
    \Else{\Return{$\fals$}}
  }
\vspace*{-1ex}
\end{algorithm}

\begin{algorithm}[t]
  \caption{\HyperPCTL model checking algorithm II}
  \label{algo:mc2}
  \SetKwFunction{HyperPCTL}{HyperPCTL}
  \Input{DTMC $\dtmc=(S, \tpm, \AP, L)$, \HyperPCTL quantified formula $\psi$, 
non-negative integer $n$, $\hat{L}^n:S^n\rightarrow 2^\mathcal{F}$}
  \Output{An extension of $\hat{L}^n$ to label each state $s\in S^n$ with 
sub-formulas of $\psi$ that hold in $s$}
  \Fn{HyperPCTL{($\dtmc,\psi,n,\hat{L}^n$)}}{
    \If{$\psi=\tru$}{for all $s\in S^n$ set $\hat{L}^n(s) := 
\hat{L}^n(s)\cup\{\tru\}$} \label{line:true}
    \ElseIf{$\psi=a_{\sigma_{i}}$}{for all $s\in S^n$ with $a_{i}\in L^n(s)$ set 
$\hat{L}^n(s):=\hat{L}^n(s)\cup\{a_{\sigma_{i}}\}$} \label{line:prop}
    \ElseIf{$\psi=\psi_1\wedge\psi_2$\label{line:and1}}{
      $\hat{L}^n$:=HyperPCTL($\dtmc$,$\psi_1$,$n$,$\hat{L}^n$)\;
      $\hat{L}^n$:=HyperPCTL($\dtmc$,$\psi_2$,$n$,$\hat{L}^n$)\;      
      for all states $s\in S^n$ with $\{\psi_1,\psi_2\}\subseteq \hat{L}^n(s)$ 
set $\hat{L}^n(s):=\hat{L}^n(s)\cup\{\psi\}$\label{line:and2}} 
    \ElseIf{$\psi=\neg\psi_1$ \label{line:neg1}}{
      $\hat{L}^n$:=HyperPCTL($\dtmc$,$\psi_1$,$n$,$\hat{L}^n$)\;      
      for all states $s\in S^n$ with $\psi_1\notin\hat{L}^n(s)$ set 
$\hat{L}^n(s):=\hat{L}^n(s)\cup\{\psi\}$ \label{line:neg2}
    }
    \ElseIf{$\psi=p_1\sim p_2$\label{line:pctl1}}{
      $L^n_1$ := ProbMC($\dtmc,p_1,n,\hat{L}^n$) \hfill {\em \%  see
Algorithm~\ref{algo:mc3}}\;
      $L^n_2$ := ProbMC($\dtmc,p_2,n,\hat{L}^n$) \hfill {\em \%  see
Algorithm~\ref{algo:mc3}}\;
      for all states $s\in S^n$ with $L^n_1(s)\sim L^n_2(s)$ set 
$\hat{L}^n(s):=\hat{L}^n(s)\cup\{\psi\}$\label{line:pctl2}
    }
    \ElseIf{$\psi=\exists\sigma_i.\psi_1$\label{line:exists1}}{
      $\hat{L}^n$:=HyperPCTL($\dtmc$,$\psi_1$,$n$,$\hat{L}^n$)\;
      for all states $s=(s_1,\ldots,s_n)\in S^n$ with $\psi_1\in\hat{L}^n(s')$ 
for some $s_i'\in S$ and $s'=(s_1,\ldots,s_{i-1},s_i',s_{i+1},\ldots,s_n)$ set 
$\hat{L}^n(s):=\hat{L}^n(s)\cup\{\psi\}$\label{line:exists2}
    }
    \ElseIf{$\psi=\forall\sigma_i.\psi_1$\label{line:forall1}}{
      $\hat{L}^n$:=HyperPCTL($\dtmc$,$\psi_1$,$n$,$\hat{L}^n$)\;
      for all states $s=(s_1,\ldots,s_n)\in S^n$ with $\psi_1\in\hat{L}^n(s')$ 
for all $s_i'\in S$ and $s'=(s_1,\ldots,s_{i-1},s_i',s_{i+1},\ldots,s_n)$ set 
$\hat{L}^n(s):=\hat{L}^n(s)\cup\{\psi\}$ \label{line:forall2}
    }
    \Return{$\hat{L}^n$}
  }
\vspace*{-1ex}
\end{algorithm}

\begin{algorithm}[t]
  \caption{\textsf{\small HyperPCTL} model checking algorithm III}
  \label{algo:mc3}
  \SetKwFunction{ProbMC}{ProbMC}
  \Input{DTMC $\dtmc=(S, \tpm, \AP, L)$, \HyperPCTL probability expression $p$, 
non-negative integer $n$, $\hat{L}^n:S^n\rightarrow 2^\mathcal{F}$}
  \Output{$L^n_p:S^n\rightarrow \mathbb{Q}$ specifying the values $L^n_p(s)$ of 
$p$ in all states $s\in S^n$}
  \Fn{ProbMC{$(\dtmc,p,n,\hat{L}^n)$}}{
    let $L^n_p:S^n\rightarrow \mathbb{Q}$ with $L^n_p(s)=0$ for all $s\in S^n$\;
    \If{$p=c$}{
      for all $s\in S^n$ set $L^n_p(s)=c$
    }
    \ElseIf{$p=p_1 \textit{ op } p_2$ with $\textit{op}\in\{+,-,\cdot\}$}{
      $L^n_1$ := probMC($\dtmc,p_1,n,\hat{L}^n$)\;
      $L^n_2$ := probMC($\dtmc,p_2,n,\hat{L}^n$)\;
      for each $s\in S^n$ set $L^n_p(s) := L^n_1(s) \textit{ op } L^n_2(s)$
    }
    \ElseIf{$p=\pr(\varphi)$}{
      \If{$\varphi=\X\psi$}{
        for all $s\in S^n$ set $L^n_p(s)=\sum_{s'\in S^n,\ \psi\in\hat{L}^n(s')} 
\tpm(s,s')$
      }
      \ElseIf{$\varphi=\psi_1\U\psi_2$}{
        compute the unique solution $\nu$ for the following equation system:\;
        (1) $p_s=0$ for all states $s\in S^n$ with $\psi_1\notin\hat{L}^n(s)$ 
and $\psi_2\notin\hat{L}^n(s)$, or if no state $s'$ with 
$\psi_2\in\hat{L}^n(s')$ is reachable from $s$\;
        (2) $p_s=1$ for all states $s\in S^n$ with $\psi_2\in\hat{L}^n(s)$\;
        (3) $p_s=\sum_{s'\in S^n} \tpm(s,s')\cdot p_{s'}$ for all other states\;
        for all $s\in S^n$ set $L^n_p(s)=\nu(p_s)$
      }
      \ElseIf{$\varphi=\psi_1\U^{[k_1,k_2]}\psi_2$}{
        for each $s\in S^n$ set $P^n_0(s)=1$ if $\psi_2\in\hat{L}^n(s)$ and 
$P^n_0(s)=0$ otherwise\;
        \For{$i=1$ to $k_2$}{
          for each $s\in S^n$ set $P^n_i(s)=\sum_{s'\in S^n}\tpm(s,s')\cdot 
P^n_{i-1}(s')$ if $\psi_1\in\hat{L}^n(s)$ and $P^n_i(s)=0$ otherwise\;
        }
        for all $s\in S^n$ set $L^n_p(s)=\sum_{i=k_1}^{k_2}P^n_i(s)$
      }
    }
    \Return{$L^n_p$}
  }
\vspace*{-1ex}
\end{algorithm}

\begin{theorem}
\label{thrm:poly}
For a finite Markov chain $\dtmc$ and \HyperPCTL formula $\psi$, the 
\HyperPCTL model checking problem, i.e., $\dtmc \models \psi$ can be solved 
in time $O(\mathsf{poly}(|\dtmc|))$.
\end{theorem}

% \begin{proof}
%  \todo{TBD}
% \end{proof}

\begin{theorem}
\label{thrm:pspace}
The \HyperPCTL model checking problem is \comp{PSPACE-hard} in the number of quantifiers in the 
formula.
\end{theorem}

%\begin{proof}
\smallskip\noindent\emph{Proof.}\  
We show that the \HyperPCTL model checking problem is \comp{PSPACE-hard} by reducing the
following \comp{PSPACE}-hard {\em quantified Boolean formula} (QBF) satisfiability problem~\cite{gj79} to it:
\begin{quote}
\emph{Given is a set $\{x_1, x_2, \dots, x_n\}$ of Boolean variables and a 
quantified Boolean formula
$$y=\quant_1 x_1.\quant_1 x_2\dots\quant_{n-1} x_{n-1}.\quant_n x_n.\psi$$
where $\quant_i \in \{\forall, \exists\}$  for each $i \in [1, n]$ and $\psi$ is
an arbitrary Boolean formula over variables $\{x_1, \ldots, x_n\}$. Is $y$ 
true?}
\end{quote}

\begin{wrapfigure}{r}{3cm}
 \centering
\vspace*{-6ex}
\begin{tikzpicture}

\node[draw,circle,text width=0.5cm] (s0) at (0,0) {};
\node at (0,0) {$s_0$};
\node[draw,circle,text width=0.5cm] (s1) at (1,0) {};
\node at (1,0) {$s_1$};
%\node[draw,circle,text width=0.5cm] (s2) at (0,-1.3) {};
%\node at (0,-1.3) {$s_2$};
%\node[draw,circle,text width=0.5cm] (s3) at (2,-1.3) {};
%\node at (2,-1.3) {$s_3$};

\node[left of=s0,node distance=0.7cm] {$\{x\}$};
\node[right of=s1,node distance=0.7cm] {$\emptyset$};
%\node[right of=s2,node distance=0.7cm] {$\{x\}$};

%\path[draw,-latex',thick] (s0) edge node[left,font=\scriptsize,near start] {$1$} (s2);
%\path[draw,-latex',thick] (s1) edge node[right,font=\scriptsize,near start] {$1$} (s3);
\path[draw,-latex',thick] (s0) edge [loop below] node[right,font=\scriptsize] {$1$} (s0);
\path[draw,-latex',thick] (s1) edge [loop below] node[right,font=\scriptsize] {$1$} (s1);
\end{tikzpicture}
\vspace*{-2ex}
 \caption{DTMC in the proof of Thm~\ref{thrm:pspace}.}
 \label{fig:pspace}
\vspace*{-3ex}
\end{wrapfigure}
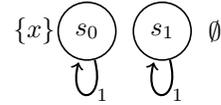

We reduce the satisfiability problem for a quantified Boolean formula to the model 
checking problem for a \HyperPCTL formula with the same quantifier structure as follows. We define the simple DTMC $\dtmc = (S, \tpm, \AP, L)$ shown in
  Fig.~\ref{fig:pspace}, which contains two states $s_0$ and $s_1$ and
  has two paths $s_0^\omega$ and $s_1^\omega$.
The \HyperPCTL formula in our mapping is the 
following:
\begin{equation}
\label{eq:hpctl-com}
 \quant_1 \sigma_{1}.\quant_1 \sigma_{2}\dots\quant_{n-1} 
\sigma_{{n-1}}.\quant_n 
\sigma_n.\, \psi'
\end{equation}
where $\psi'$ is constructed from $\psi$ by replacing every occurrence
of a variable $x_i$ in $\psi$ 
by $x_{\sigma_i}$. The given quantified Boolean formula is $\tru$ if and only if the DTMC obtained by our mapping 
satisfies HyperPCTL formula~(\ref{eq:hpctl-com}). We translate every assignment 
to the trace quantifiers to a corresponding assignment of the Boolean variables, 
and vice versa, as follows: Assigning state $s_0$ ($s_1$) to $\sigma_i$ means that 
$x_i$ is set to $\tru$ ($\fals$).
\hfill\qed

\medskip

%\end{proof}

\section{Related Work}
\label{sec:related}

{\em Probabilistic noninterference}~\cite{gray90,g92} establishes connection 
between information theory and information flow by employing probabilities to 
address covert channels. Intuitively, it requires that the probability of every 
pattern of low-observable trace be the same for every low-equivalent initial 
state. Most efforts in reasoning about probabilistic noninterference is through 
probabilistic weak bisimulation (e.g.~\cite{smith03}). More recently, Sabelfeld 
and Sands~\cite{ss00} introduce a framework to ensure nonintereference for 
multi-threaded programs, where a probabilistic scheduler non-deterministically 
manages the execution of threads. They introduce operational semantics for a 
simple imperative language with dynamic thread creation, and how 
compositionality is ensured.

{\em Epistemic logic}~\cite{fhmv95} is a subfield of modal logic that is 
concerned with reasoning about knowledge. The semantic model of the logic is a 
Kripke structure, where a set of agents are related with each other based on 
which states they consider possible. A probabilistic version of the 
logic~\cite{ht89} assigns a probability function to each agent at each state 
such that its domain is a non-empty subset of the set of possible states. 
Epistemic temporal logic has been used to express information-flow security 
policies (e.g.,~\cite{bdg11}). The relation between the expressive power of 
probabilistic epistemic logic and \HyperPCTL remains an open question in this 
paper. Gray and Syverson~\cite{gs98} propose a modal logic for multi-level 
reasoning about security of probabilistic systems. The logic is axiomatic and 
is based on Halpern and Tuttle~\cite{ht89} framework for reasoning about 
knowledge and probability. The logic is sound, but it may run into 
undecidability.

Clarkson and Schneider~\cite{cs10} introduce the notion of {\em 
hyperproperties}, a set-theoretic framework for expressing security policies. A 
hyperproperty is a set of sets of traces. In other words, a hyperproperty is a 
second-order property of properties. The expressive power of hyperproperties do 
not exceed the second-order logic, but it is currently unclear whether the full 
power of second-order logic is needed to express hyperproperties of interest. 
Clarkson and Schneider have shown two fundamental things: (1) a 
hyperproperty is an intersection of a safety and a liveness hyperproperty, 
and (2) hyperproperties can express many important requirements such as 
information-flow security policies (e.g., nonintereference, observational 
determinism, etc), service-level agreement, etc. 

Second-order logic is not verifiable in general, as it cannot be effectively 
and completely axiomatized. Thus, temporal logics for subclasses of 
hyperproperties have emerged~\cite{cfkmrs14}. \HyperLTL and \HyperCTLstar allow 
explicit and simultaneous quantification over multiple paths to \LTL and to 
\CTLstar, respectively. As the names suggest, \HyperLTL allow quantification of 
linear traces and \HyperCTLstar permits quantification over multiple execution 
traces simultaneously while allowing branching-time paths for each trace. 
\HyperLTL and \HyperCTLstar are not equipped with probabilistic 
operators and cannot reason about probabilistic systems.

% talk about $k$-safety. a program property is said to be a
% k-safety property if it can be refuted by observing k number of (finite) 
% execution traces.

\section{Conclusion and Future Work}
\label{sec:concl}

% More sophisticated formulas (e.g., entropy-based QIF)
% $$\varphi_{\mathsf{qif}} = \neg \exists \sigma_1. \dots .\exists_{\log |H|}. 
% \sum_{i=1}^{\log |L|} \pr(\F l) \times \sum_{j=0}^{\log |H|}\bigg[\pr(h_\sigma 
% | l_\sigma) \times \log \frac{1}{\pr (h_\sigma | l_\sigma)}\bigg] \leq q$$

In this paper, we proposed the temporal logic \HyperPCTL to express and reason 
about probabilistic hyperproperties. \HyperPCTL is a natural extension to 
\PCTL by allowing explicit and simultaneous quantification over model states. We defined the syntax and semantics and presented a 
model checking algorithm for discrete-time Markov chains. The complexity of the 
algorithm is 
%polynomial time in the size of the input Markov chain and is 
\comp{PSPACE-hard} in the number of quantifiers in the input \HyperPCTL formula. We
presented multiple examples from different domains, where \HyperPCTL can 
elegantly express the requirements. 
%These examples were probabilistic 
%noninterference, quantitative information flow, differential privacy, 
%probabilistic causation, and software reliability testing.

We believe the results in this paper 
%open numerous open problems and 
paves 
the path for new research directions. As for future work, an important 
unanswered question in this paper is to determine tighter lower and upper bounds 
for the the complexity of \HyperPCTL model checking in the size of the formula. 
We believe most of the literature and fundamental lines of research on \PCTL 
verification should now be revisited in the context of \HyperPCTL. Examples 
include \HyperPCTL model checking for Markov decision processes (MDPs), Markov 
chains with costs, parameter synthesis and model repair for probabilistic 
hyperproperties, \HyperPCTL conditional probabilities, developing 
abstraction/refinement, comparing expressive power to existing related
logics such as probabilistic epistemic logic~\cite{ht89}, etc. An orthogonal 
direction is deeper investigation of the examples presented in 
Section~\ref{sec:examples}. Each of those areas (e.g., differential privacy and 
probabilistic causation) deserve more research to develop effective and 
efficient model checking techniques.

\section{Acknowledgments}

We thank Boris K\"{o}pf for his valuable insights on expressing 
QIF policies. 

\bibliographystyle{plain}
\bibliography{bibliography}

\end{document}